\documentclass[a4paper,10pt]{article}
\usepackage{stmaryrd}
\usepackage{amsfonts}
\usepackage{bbm}
\usepackage{amscd}
\usepackage{mathrsfs}
\usepackage{latexsym,amssymb,amsmath,amscd,amscd,amsthm,amsxtra}
\usepackage[dvips]{graphicx}
\usepackage[utf8]{inputenc}
\usepackage[T1]{fontenc}
\usepackage{lmodern}
\usepackage{amssymb}
\usepackage[all]{xy}
\usepackage{nicefrac,mathtools,enumitem}
\usepackage{microtype}
\usepackage{xcolor}
\newtheorem{thm}{Theorem}[section]
\newtheorem{lem}[thm]{Lemma}
\newtheorem{cor}[thm]{Corollary}
\newtheorem{pro}[thm]{Proposition}

\newtheorem{defi}[thm]{Definition}

\newcommand {\emptycomment}[1]{}

\newcommand{\be }{\begin{equation}}
\newcommand{\ee }{\end{equation}}

\newcommand{\pf}{\noindent{\bf Proof.}\ }

\newcommand{\Nat}{\mathbb N}

\newcommand{\huaR}{\mathcal{R}}

\newcommand{\huaO}{\mathcal{O}}

\newcommand{\g}{\mathfrak g}

\def\qed{\hfill ~\vrule height6pt width6pt depth0pt}

\newcommand{\Courant}[1]{\left\llbracket  #1\right\rrbracket }

\newcommand{\Id}{\rm{Id}}

\newcommand{\br}[1]{   [ \cdot,    \cdot  ]   }

\newcommand{\gl}{\mathfrak {gl}}

\newcommand{\ad}{\mathrm{ad}}

\begin{document}
\title {{Kupershmidt-(dual-)Nijenhuis structures on a Lie algebra with a representation}
 }\vspace{2mm}
 \author{Yuwang Hu$^1$, Jiefeng Liu$^1$ and Yunhe Sheng$^{2}$  \\
$^1$School of Mathematics and Statistics, Xinyang Normal University,\\ Xinyang 464000, Henan, China\\
$^2$Department of Mathematics, Jilin University,\\ Changchun 130012, Jilin, China
\\\vspace{3mm}
Email: hywzrn@163.com,~liujf12@126.com,~shengyh@jlu.edu.cn }
\date{}
\maketitle

\begin{abstract}
In this paper, first we study infinitesimal deformations of a Lie algebra with a representation and introduce the notion of a Nijenhuis pair, which gives a trivial deformation of a Lie algebra with a representation. Then we introduce the notion of a Kupershmidt-(dual-)Nijenhuis structure  on a Lie algebra with a representation, which is a generalization of the $r$-$n$ structure ($r$-matrix-Nijenhuis structure) introduced by Ravanpak,   Rezaei-Aghdam and   Haghighatdoost. We show that a Kupershmidt-(dual-)Nijenhuis structure gives rise to a hierarchy of Kupershmidt operators. Finally, we define a Rota-Baxter-Nijenhuis structure to be a  Kupershmidt-Nijenhuis structure  on a Lie algebra with respect to the adjoint representation, and study the relation between Rota-Baxter-Nijenhuis structures and $r$-matrix-Nijenhuis structures.\\\\
{\bf Keywords}: (dual-)Nijenhuis pair, Kupershmidt operator, Kupershmidt-(dual-)Nijenhuis structure, $r$-matrix-Nijenhuis structure, Rota-Baxter-Nijenhuis structure\\
{\bf MSC}: 17B60, 17B99, 17D99
\end{abstract}
\section{Introduction}
Nijenhuis operators on Lie algebras have been introduced in the theory of integrable systems in the work of Magri, Gelfand and Dorfman (see the book \cite{Dorfman}), and, under the name of hereditary operators, in that of Fuchssteiner and Fokas (\cite{FuchFoka}). In the sense of the theory of deformations of Lie algebras (\cite{Nijenhuis}), Nijenhuis operators generate trivial deformations naturally. More precisely,  A {\bf Nijenhuis operator} on a Lie algebra $(\g,[-,-]_\g)$ is a linear map $N:\g\longrightarrow\g$ satisfying
\begin{equation}
  [N(x),N(y)]_\g=N\big([N(x),y]_\g+[x,N(y)]_\g-N[x,y]_\g\big),\quad \forall~x,y\in\g.
\end{equation}
Furthermore, the deformed bracket $[-,-]_N:\wedge^2\g\longrightarrow\g$ given by
\begin{equation}\label{eq:deformbracket}
  [x,y]_N=[N(x),y]_\g+[x,N(y)]_\g-N([x,y]_\g),
\end{equation}
 is a Lie bracket and $N$ is a Lie algebra morphism from $(\g,[-,-]_N)$ to $(\g,[-,-]_\g)$.

 Poisson-Nijenhuis structures were defined by Magri and Morosi in 1984 (\cite{MaMo}) in their study of completely integrable systems. See \cite{Kosmann1,Kosmann2} for more details on Poisson-Nijenhuis structures. Recently, Ravanpak,   Rezaei-Aghdam and   Haghighatdoost introduced the notion of an $r$-$n$ structure on a Lie algebra $\g$, which is the infinitesimal of a right-invariant Poisson-Nijenhuis structure on the  Lie group $G$ integrating the Lie algebra $\g$ (\cite{rn}).  An   $r$-$n$ structure on a Lie algebra $\g$ is a pair $(\pi,N)$, where $\pi$ is an $r$-matrix, i.e. $\pi$ satisfies the classical Yang-Baxter equation $[\pi,\pi]_\g=0$,   and $N$ is a Nijenhuis operator on $\g$, such that some compatibility conditions are satisfied. In this paper, we will call such a structure an   $r$-matrix-Nijenhuis structure. An equivalent description of a bivector   $\pi\in\wedge^2\g$ being an  $r$-matrix is given by
\begin{equation}\label{eq:r-matrix}
  {[\pi^\sharp(\alpha),\pi^\sharp(\beta)]}_\g=\pi^\sharp(\ad^*_{\pi^\sharp(\alpha)}\beta-\ad^*_{\pi^\sharp(\beta)}\alpha),\quad\forall~\alpha,\beta\in\g^*,
\end{equation}
where   $\pi^\sharp:\g^*\longrightarrow\g$ is defined  by
$\langle\pi^\sharp(\alpha),\beta\rangle=\pi(\alpha,\beta)$ and
 $\ad^*$ is the coadjoint representation of the Lie algebra $\g$. See \cite{Drinfeld,Kosmann0} for more details.

 Baxter introduced the concept of a Rota-Baxter algebra  for associative algebras   (\cite{Ba}) in his study of fluctuation theory in probability. It has been found many applications in recent years, including  the algebraic approach of Connes-Kreimer~(\cite{CK}) to renormalization of perturbative quantum field theory, tridendriform
algebras (\cite{BGN2013}), quantum analogue of Poisson geometry (\cite{Uchino1}), twisting on associative algebras (\cite{Uchino2}). In the Lie algebra context, a Rota-Baxter operator of weight zero was introduced independently in the 1980s as the operator form of the classical Yang-Baxter equation, named after the physicists C.-N. Yang and R. Baxter. See the book \cite{Gub} for more details. A linear operator $\huaR:\g\longrightarrow\g$ on a Lie algebra $\g$   is called a Rota-Baxter operator if the following condition is satisfied:
\begin{equation}\label{eq:Rota-Baxter}
  [\huaR(x),\huaR(y)]_\g=\huaR([\huaR(x),y]_\g+[x,\huaR(y)]_\g),\quad\forall~x,y\in\g.
\end{equation}
In \cite{Semonov-Tian-Shansky}, Semonov-Tian-Shansky studied the classical Yang-Baxter equation systematically.
He proved that if there is an $\ad$-invariant, non-degenerate, symmetric bilinear form on $\g$, then a Rota-Baxter operator and an $r$-matrix are equivalent.   Moreover, Kupershmidt restudied the classical Yang-Baxter equation (\cite{Kuper1}) and generalized the above representation to   an arbitrary representation $(V;\rho)$ of $\g$, and introduce the notion of a Kupershmidt operator (also called an $\huaO$-operator). By definition, a Kupershmidt operator on a Lie algebra $\g$ with respect to a representation $(V;\rho)$ is a
linear map $T:V\longrightarrow \g$ satisfying
\begin{equation}
  [T(u), T(v)]_\g=T\Big(\rho(T(u))(v)-\rho(T(v))(u)\Big),\quad \forall~u,v\in V.
\end{equation}
  Note that a Rota-Baxter operator given by  Semonov-Tian-Shansky is just a Kupershmidt operator on a Lie algebra with respect  to the adjoint representation $(\g;\ad)$ and an $r$-matrix is a Kupershmidt operator on a Lie algebra with respect to the coadjoint representation $(\g^*;\ad^*)$. Moreover, the notion of an extended $\huaO$-operator was introduced by Bai, Guo and Ni in \cite{BGN2010,BGN2011} and plays important role in the study of nonabelian generalized Lax pairs and the extended classical Yang-Baxter equation.

The purpose of this paper is to introduce the notion of a Kupershmidt-(dual-)Nijenhuis structure on a Lie algebra with respect to a representation that contains the aforementioned $r$-matrix-Nijenhuis structure as a special case, and give applications.

First we study   infinitesimal deformations of a Lie algebra with a representation and introduce the notion of a Nijenhuis pair. Similar to that a Nijenhuis operator gives a trivial infinitesimal deformation of a Lie algebra, a Nijenhuis pair  gives a trivial infinitesimal deformation of a Lie algebra with a representation. We also introduce the notion of a dual-Nijenhuis pair, which can be viewed as the dual of a Nijenhuis pair. Based on the Nijenhuis pair and dual-Nijenhuis pair, we introduce the notions of a Kupershmidt-Nijenhuis structure and a Kupershmidt-dual-Nijenhuis structure, respectively.
Just as that a Poisson-Nijenhuis structure gives rise to a hierarchy  of Poisson structures, a Kupershmidt-(dual-)Nijenhuis structure also gives rise to a hierarchy of Kupershmidt operators and these Kupershmidt operators are pairwise compatible. Moreover,  compatible Kupershmidt operators can also give rise to a Kupershmidt-dual-Nijenhuis structure. Our definition of a Kupershmidt-dual-Nijenhuis structure satisfies the criterion that a Kupershmidt-dual-Nijenhuis structure on a Lie algebra with respect to the coadjoint representation $(\g^*;\ad^*)$ is exactly an $r$-matrix-Nijenhuis structure.  Since a Rota-Baxter operator is a Kupershmidt operator with respect to the adjoint representation, we define a Rota-Baxter-Nijenhuis structure to be a Kupershmidt-Nijenhuis structure on a Lie algebra with respect to the adjoint representation $(\g;\ad)$. The relation between  Rota-Baxter-Nijenhuis structure and $r$-matrix-Nijenhuis structure is investigated (see Theorem \ref{thm:rmatrix-RBN}).

\emptycomment{ From the theory of the Kupershmidt-(dual-)Nijenhuis structures, the main results of $r$-matrix-Nijenhuis structures have been absorbed into this general framework.Therefore, once there is a structure satisfying the conditions of Kupershmidt-(dual-)Nijenhuis structures, an analogue of the results of $r$-matrix-Nijenhuis structures follow immediately. We give an example of Kupershmidt-dual-Nijenhuis structure through pseudo-Hessian-Nijenhuis structures on a pre-Lie algebra, which is introduced in \cite{WBLS} similar to the symplectic-Nijenhuis structures on a Lie algebra.
Furthermore, there is an important algebraic structure behind the Kupershmidt operators. It is the pre-Lie algebra (also called left-symmetric algebras, quasi-associative algebras, Vinberg algebras and so on). Pre-Lie algebras are a class of nonassociative algebras coming from the study of convex homogeneous cones, affine manifolds and affine structures on Lie groups, deformation of associative algebras and then  appeared in many fields in mathematics and mathematical physics, such as complex and symplectic structures on Lie groups and Lie algebras, integrable systems, Poisson brackets and infinite dimensional Lie algebras, vertex algebras, quantum field theory, operads and so on. See \cite{Andrada,Bakalov,ChaLiv,Lichnerowicz}, and the survey \cite{Pre-lie algebra in geometry} and the references therein for more details.
}

The paper is organized as follows. In Section 2, we study   infinitesimal deformations of a Lie algebra with a representation, and introduce the notion of a Nijenhuis pair on a Lie algebra and show that it generates a trivial deformation of a Lie algebra with a representation.  We also introduce the notion of a dual-Nijenhuis pair as the dual of a Nijenhuis pair. In Section 3, we introduce the notions of a Kupershmidt-Nijenhuis structure and a Kupershmidt-dual-Nijenhuis structure. Some properties of Kupershmidt-(dual-)Nijenhuis structures are studied. In Section 4, we first give the relations between Nijenhuis operators and Kupershmidt operators. Then, we prove that, on the one hand, Kupershmidt-(dual-)Nijenhuis structures give rise to hierarchies of Kupershmidt operators, which are pairwise compatible; on the other hand,  compatible Kupershmidt operators with a condition can give a Kupershmidt-dual-Nijenhuis structure. In Section 5, we introduce the notion of a Rota-Baxter-Nijenhuis structure on a Lie algebra and prove that it is equivalent to the $r$-matrix-Nijenhuis structure with some conditions. 
\section{Infinitesimal deformations of a Lie algebra with a representation}

Let $(\g,[-,-]_\g)$ be a Lie algebra and $\rho:\g\longrightarrow\gl(V)$ a representation. Let $\omega:\wedge^2\g\longrightarrow\g$ and $\varpi:\g\longrightarrow\gl(V)$ be linear maps. Consider a $t$-parametrized family of bracket operations and linear maps:
\begin{eqnarray*}
  [x,y]_t&=&[x,y]_\g+t\omega(x,y),\\
  \rho_t(x)&=&\rho(x)+t\varpi(x).
\end{eqnarray*}
 If $(\g,[-,-]_t)$ are Lie algebras and $\rho_t$ are representations of $(\g,[-,-]_t)$ on $V$ for all $t$, we say that $(\omega,\varpi)$ generates a one-parameter infinitesimal deformation of the Lie algebra  $(\g,[-,-]_\g)$ with the representation $(V;\rho)$. We denote a one-parameter infinitesimal deformation of a Lie algebra  $(\g,[-,-]_\g)$ with a representation $(V;\rho)$ by $(\g,[-,-]_t,\rho_t)$.

By direct calculation, we can deduce that $(\g,[-,-]_t,\rho_t)$ is a one-parameter infinitesimal deformation of the Lie algebra  $(\g,[-,-]_\g)$ with the representation $(V;\rho)$ if and only if
\begin{eqnarray}
 \label{eq:deform1}[\omega(x,y),z]_\g+ [\omega(z,x),y]_\g+ [\omega(y,z),x]_\g&=&\omega(x,[y,z]_\g)++\omega(z,[x,y]_\g)+\omega(y,[z,x]_\g),\\
 \label{eq:deform2}\omega(\omega(x,y),z)+\omega(\omega(z,x),y)+\omega(\omega(y,z),x)&=&0,\\
 \label{eq:deform3}\varpi(\omega(x,y))&=&[\varpi(x),\varpi(y)],\\
 \label{eq:deform4}\rho(\omega(x,y))+\varpi([x,y]_\g)&=&[\rho(x),\varpi(y)]+[\varpi(x),\rho(y)].
\end{eqnarray}
It is well-known that \eqref{eq:deform1} means that $\omega$ is a $2$-cocycle of the Lie algebra $\g$ with the coefficient in the adjoint representation and \eqref{eq:deform2} means that $(\g,\omega)$ is a Lie algebra. Furthermore, \eqref{eq:deform3} means that $\varpi$ is a representation of the Lie algebra $(\g,\omega)$ on $V$ and \eqref{eq:deform4} means that $\rho+\varpi$ is a representation of the Lie algebra $(\g,[-,-]_\g+\omega(-,-))$ on $V$.

 \begin{defi}Two one-parameter infinitesimal deformations, $(\g,[-,-]_t,\rho_t)$ and $(\g,[-,-]'_t,\rho'_t)$,  of a Lie algebra  $(\g,[-,-]_\g)$ with a representation $(V;\rho)$ are equivalent if there exists an isomorphism $({\Id}_\g+tN,{\Id}_V+tS)$ from $(\g,[-,-]'_t,\rho'_t)$ to $(\g,[-,-]_t,\rho_t)$, i.e.
 \begin{eqnarray*}
   ({\Id}_\g+tN)[x,y]'_t&=&[({\Id}_\g+tN)(x),({\Id}_\g+tN)(y)]_t,\\
   \quad ({\Id}_V+tS)\circ\rho'_t(x)&=&\rho_t(({\Id}_\g+tN)(x))\circ ({\Id}_V+tS).
 \end{eqnarray*}

   A one-parameter infinitesimal deformation  of a Lie algebra  $(\g,[-,-]_\g)$ with a representation $(V;\rho)$ is said to be {\bf trivial} if it is equivalent to $(\g,[-,-]_\g,\rho)$.
 \end{defi}
By straightforward computations, $(\g,[-,-]_t,\rho_t)$ is a trivial deformation if and only if
\begin{eqnarray}
\label{eq:trivialdeform1}\omega(x,y)&=&[N(x),y]_\g+[x,N(y)]_\g-N([x,y]_\g),\\
\label{eq:trivialdeform2}N\omega(x,y)&=&[N(x),N(y)]_\g,\\
\label{eq:trivialdeform3}\varpi(x)&=&\rho(N(x))+\rho(x)\circ S-S\circ\rho(x),\\
\label{eq:trivialdeform4}\rho(N(x))\circ S&=&S\circ \varpi(x).
\end{eqnarray}
It follows from \eqref{eq:trivialdeform1} and \eqref{eq:trivialdeform2} that $N$ must be a Nijenhuis operator on the Lie algebra $(\g,[-,-]_\g)$. It follows from \eqref{eq:trivialdeform3} and \eqref{eq:trivialdeform4} that $N$ and $S$ should satisfy the condition:
\begin{equation}\label{eq:Nijpair}
     \rho(N(x))(S(v))=S(\rho(Nx)(v))+S(\rho(x)S(v))-S^2(\rho(x)(v)).
   \end{equation}
 \begin{defi}
   A pair $(N,S)$, where $N\in\gl(\g)$ and $S\in\gl(V)$, is called a {\bf Nijenhuis pair} on a Lie algebra  $(\g,[-,-]_\g)$ with a representation $(V;\rho)$ if $N$ is a Nijenhuis operator on the Lie algebra $(\g,[-,-]_\g)$ and the condition \eqref{eq:Nijpair} holds.
 \end{defi}
We have seen that a trivial deformation of a Lie algebra with a representation could give rise to a Nijenhuis pair. In fact, the converse is also true.
\begin{thm}
  Let $(N,S)$ be a Nijenhuis pair on a Lie algebra  $(\g,[-,-]_\g)$ with a representation $(V;\rho)$. Then a deformation of $(\g,[-,-]_\g,\rho)$ can be obtained by putting
  \begin{eqnarray}
    \omega(x,y)&=&[N(x),y]_\g+[x,N(y)]_\g-N([x,y]_\g);\\
    \varpi(x)&=&\rho(N(x))+\rho(x)\circ S-S\circ\rho(x).
  \end{eqnarray}
  Furthermore, this deformation is trivial.
\end{thm}
\pf It is a straightforward computations. We omit the details.\qed
\vspace{3mm}

Similar to the definition of a Nijenhuis pair, we introduce the notion of a dual-Nijenhuis pair on a Lie algebra with a representation.
\begin{defi}
A pair $(N,S)$, where $N\in\gl(\g)$ and $S\in\gl(V)$, is called a {\bf dual-Nijenhuis pair} on a Lie algebra  $(\g,[-,-]_\g)$ with a representation $(V;\rho)$ if $N$ is a Nijenhuis operator on the Lie algebra $(\g,[-,-]_\g)$ and $S$ satisfies the following condition:
   \begin{equation}\label{eq:coNijpair}
    \rho({N(x)})(S(v))=S(\rho({N(x)})(v))+\rho({x})(S^2(v))-S(\rho({ x})(S(v))).
   \end{equation}
 \end{defi}
Let $\rho^*:\g\longrightarrow \gl(V^*)$ be the dual representation of $\rho$ defined by
\begin{eqnarray}
 \langle \rho^*(x)(\alpha),v\rangle=-\langle \alpha,\rho(x)(v)\rangle,\quad\forall~x,y\in\g,v\in V,\alpha\in V^*.
\end{eqnarray}
In fact, there is a close relationship between a Nijenhuis pair and a dual-Nijenhuis pair.
\begin{pro}\label{pro:Nij-coNij pair}
$(N,S)$ is a Nijenhuis pair on a Lie algebra  $(\g,[-,-]_\g)$ with a representation $(V;\rho)$ if and only if $(N,S^*)$ is a dual-Nijenhuis pair on the Lie algebra  $(\g,[-,-]_\g)$ with the representation $(V^*;\rho^*)$.
\end{pro}
\pf It follows from
\begin{eqnarray*}
   &&\langle\rho(N(x))S(v)-S\rho(N(x))(v)-S\rho(x)S(v)+S^2\rho(x)(v),\alpha\rangle\\
   &=&-\langle v,S^*\rho^*(N(x))(\alpha)-\rho^*(N(x))(S^*(\alpha))-S^*\rho^*(x)(S^*(\alpha))+\rho^*(x)((S^*)^2(\alpha))\rangle.\qed
\end{eqnarray*}

\begin{defi}
  A Nijenhuis pair $(N,S)$ on a Lie algebra  $(\g,[-,-]_\g)$ with a representation $(V;\rho)$ is called a {\bf perfect Nijenhuis pair}  if
  \begin{equation}\label{eq:pNijpair}
    S^2(\rho(x)(v))+\rho({x})(S^2(v))=2S(\rho({ x})(S(v))),\quad\forall~x\in\g,v\in V.
   \end{equation}
\end{defi}
It is obvious that a perfect Nijenhuis pair is not only a Nijenhuis pair but also a dual-Nijenhuis pair.

A Nijenhuis pair gives rise to a Nijenhuis operator on the semidirect product Lie algebra.
\begin{pro}\label{pro:bigN}
   Let $(N,S)$ be a Nijenhuis pair on a Lie algebra  $(\g,[-,-]_\g)$ with a representation $(V;\rho)$. Then $N+S$ is a Nijenhuis operator on the semidirect product Lie algebra $\g\ltimes_\rho V.$ Furthermore, if $(N,S)$ is a perfect  Nijenhuis pair, then $N+S^*$ is a also a Nijenhuis operator on the semidirect product Lie algebra $\g\ltimes_{\rho^*} V^*$.
 \end{pro}

 Define $\hat{\varrho}:\g\longrightarrow\gl(V)$ and $\tilde{\varrho}:\g\longrightarrow\gl(V)$, respectively, by
 \begin{eqnarray}
  \hat{ \varrho}(x)&=&\rho(Nx)+[\rho(x),S],\\
  \tilde{ \varrho}(x)&=&\rho(Nx)-[\rho(x),S],\quad\forall~x\in\g.
 \end{eqnarray}

 \begin{cor}\begin{itemize}
   \item[\rm(i)]If $(N,S)$ is a Nijenhuis pair on a Lie algebra  $(\g,[-,-]_\g)$ with a representation $\rho$, then $\hat{\varrho}$ is a representation of the Lie algebra $(\g,[-,-]_N)$ on $V$;
       \item[\rm(ii)]If $(N,S)$ is a dual-Nijenhuis pair on a Lie algebra  $(\g,[-,-]_\g)$ with a representation $\rho$, then $\tilde{\varrho}$ is a representation of the Lie algebra $(\g,[-,-]_N)$ on $V$;
 \end{itemize}

 \end{cor}
\pf (i) By Proposition \ref{pro:bigN}, $N+S$ is a Nijenhuis operator on the semidirect product Lie algebra  $\g\ltimes_\rho V$. The deformed bracket $[-,-]_{N+S}$ is given by
\begin{eqnarray*}
  [x+u,y+v]_{N+S}&=&[(N+S)(x+u),y+v]_\rho+[x+u,(N+S)(y+v)]_\rho-(N+S)[x+u,y+v]_\rho\\
  &=&[Nx,y]_\g+[x,Ny]_\g-N[x,y]_\g\\
  &&+\rho(Nx)(v)+\rho(x)S(v)-S\rho(x)(v)-\rho(y)S(u)-\rho(Ny)(u)+S\rho(y)(u)\\
  &=&[x,y]_N+\hat{\varrho}(x)(v)-\hat{\varrho}(y)(u),
\end{eqnarray*}
which implies that $\hat{\varrho}$ is a representation of the Lie algebra $(\g,[-,-]_N)$ on $V$.

(ii) By direct calculation, the dual map $\tilde{\varrho}^*$ of $\tilde{\varrho}$ is given by
\begin{equation}
\tilde{\varrho}^*(x)={\rho}^*(Nx)+[{\rho}^*(x),S^*],\quad\forall x\in\g.
\end{equation}
Since $(N,S)$ is a dual-Nijenhuis pair with a representation $\rho$, by Proposition \ref{pro:Nij-coNij pair}, $(N,S^*)$ is a Nijenhuis pair with a representation $\rho^*$. By (i), $\tilde{\varrho}^*$ is a representation of the Lie algebra $(\g,[-,-]_N)$ on $V^*$ and thus $\tilde{\varrho}$ is a representation of the Lie algebra $(\g,[-,-]_N)$ on $V$.\qed

\section{Kupershmidt-(dual-)Nijenhuis structures}

\emptycomment{Recall that a {\bf Kupershmidt operator} on a Lie algebra
$(\g,[-,-]_\g)$ associated to a representation $(V;\rho)$ is a
linear map $T:V\longrightarrow \g$ satisfying
\begin{equation}
  [T(u), T(v)]_\g=T\Big(\rho(T(u))(v)-\rho(T(v))(u)\Big),\quad \forall u,v\in V.
\end{equation}
If the representation is the adjoint representation $(\g;\ad)$, we obtain the usual Rota-Baxter  Lie algebra. More precisely, a Rota-Baxter  Lie algebra is a Lie algebra $(\g,[-,-]_\g)$ equipped with a linear operator $\huaR:\g\longrightarrow\g $ such that
\begin{equation}
  [\huaR(x),\huaR(y)]_\g=\huaR([\huaR(x), y]_\g+[x,\huaR(y)]_\g).
\end{equation}}

There is a close relationship between  Kupershmidt operators and pre-Lie algebras.
\begin{defi}  A {\bf pre-Lie algebra} is a pair $(V,\star)$, where $V$ is a vector space and  $\star:V\otimes V\longrightarrow V$ is a bilinear multiplication
satisfying that for all $x,y,z\in V$, the associator
$(x,y,z)=(x\star y)\star z-x\star(y\star z)$ is symmetric in $x,y$,
i.e.
$$(x,y,z)=(y,x,z),\;\;{\rm or}\;\;{\rm
equivalently,}\;\;(x\star y)\star z-x\star(y\star z)=(y\star x)\star
z-y\star(x\star z).$$
\end{defi}
Let $(V,\star)$ be a pre-Lie algebra. The commutator $
[x,y]^c=x\star y-y\star x$ defines a Lie algebra structure
on $V$, which is called the {\bf sub-adjacent Lie algebra} of
$(V,\star)$ and denoted by $V^c$. Furthermore,
$L:V\longrightarrow \gl(V)$ with $x\rightarrow L_x$, where
$L_xy=x\star y$, for all $x,y\in V$, gives a representation of
the Lie algebra $V^c$ on $V$. See \cite{Pre-lie algebra in geometry} for more details.

The following result establishes the connection between Kupershmidt operators and pre-Lie algebras.

\begin{thm}{\rm(\cite{Bai1})}
Let $T:V\to \g$ be a Kupershmidt operator  on a Lie algebra $(\g,[-,-]_\g)$ with respect to a representation $(V;\rho)$. Define a multiplication $\star^T$ on $V$ by
\begin{equation}
  u\star^T v=\rho(Tu)(v),\quad \forall u,v\in V.
\end{equation}
Then $(V,\star^T)$ is a pre-Lie algebra.
 \end{thm}

We denote by $(V,[-,-]^T)$ the sub-adjacent Lie algebra of the pre-Lie algebra $(V,\star^T)$. More precisely,
\begin{equation}\label{eq:bracketT}
 ~[u,v]^T=\rho(Tu)(v)-\rho(Tv)(u).
\end{equation}
Moreover,  $T$ is a Lie algebra homomorphism from $(V,[-,-]^T)$ to $(\g,[-,-]_\g)$.

Now let $T:V\longrightarrow\g$ be a Kupershmidt operator and $(N,S)$ a (dual-)Nijenhuis pair on a Lie algebra  $(\g,[-,-]_\g)$ with a representation $(V;\rho)$.  We define the bracket $[-,-]^T_S:\wedge^2V\longrightarrow V$ to be the deformed bracket of $[-,-]^T$ by $S$, i.e.
\begin{eqnarray*}
 {[u,v]}^T_S&=&[S(u),v]^T+[u,S(v)]^T-S([u,v]^T),\quad\forall~u,v\in V.\label{eq:defieq33}
\end{eqnarray*}
Define the bracket $\{-,-\}_{\hat{\varrho}}^T:\wedge^2V\longrightarrow V$ and $\{-,-\}_{\tilde{\varrho}}^T:\wedge^2V\longrightarrow V$ similar as \eqref{eq:bracketT} using the representation $\hat{\varrho}$ and $\tilde{\varrho}$, respectively:
\begin{eqnarray}
 \label{eq:defieq22} \{u,v\}_{\hat{\varrho}}^T&=&\hat{\varrho}(Tu)(v)-\hat{\varrho}(Tv)(u),\\
  \label{eq:defieq23}\{u,v\}_{\tilde{\varrho}}^T&=&\tilde{\varrho}(Tu)(v)-\tilde{\varrho}(Tv)(u),\quad\forall~u,v\in V.
\end{eqnarray}
It is not true in general that the brackets $[-,-]^T_S$, $\{-,-\}_{\hat{\varrho}}^T$ and  $\{-,-\}_{\tilde{\varrho}}^T$ satisfy the Jacobi identity.

\begin{defi}
  A Kupershmidt operator $T:V\longrightarrow \g$   on a Lie algebra $(\g,[-,-]_\g)$ with respect to a representation $(V;\rho)$  and a (dual-)Nijenhuis pair $(N,S)$ are called {\bf compatible} if they satisfy the following conditions
 \begin{eqnarray}
 \label{eq:TN1}N\circ T&=&T\circ S,\\
 \label{eq:TN2} {[u,v]}^{N\circ T}&=&{[u,v]}^{T}_{S}.
  \end{eqnarray}
 The triple $(T,S,N)$ is called a {\bf Kupershmidt-(dual-)Nijenhuis structure} on the Lie algebra $(\g,[-,-]_\g)$ with respect to the representation $(V;\rho)$ if $T$   and $(N,S)$ are  compatible.
 \end{defi}
Note that if $(N,S)$ is a perfect Nijenhuis pair, then a Kupershmidt-Nijenhuis structure is also a Kupershmidt-dual-Nijenhuis structure.

\begin{lem}
\begin{itemize}
\item[$\rm(a)$] Let $(T,S,N)$ be a Kupershmidt-Nijenhuis structure. Then we have
 $$ {[u,v]}^T_S=\{u,v\}_{\hat{\varrho}}^T.$$
 \item[$\rm(b)$] Let $(T,S,N)$ be a Kupershmidt-dual-Nijenhuis structure. Then we have
 $$ {[u,v]}^T_S=\{u,v\}_{\tilde{\varrho}}^T.$$
\end{itemize}
\end{lem}
\pf(a) It follows  from \eqref{eq:TN1} directly.

(b) By \eqref{eq:TN1}, we have
$${[u,v]}^T_S+\{u,v\}_{\tilde{\varrho}}^T=2{[u,v]}^{N\circ T}.$$
Then by \eqref{eq:TN2}, we obtain ${[u,v]}^T_S=\{u,v\}_{\tilde{\varrho}}^T.$\qed\vspace{3mm}

Thus, if $(T,S,N)$ is a   Kupershmidt-(dual-)Nijenhuis structure, then the three brackets  ${[-,-]}^T_S,$ $\{-,-\}_{\hat{\varrho}}^T~(\{-,-\}_{\tilde{\varrho}}^T)$ and $[-,-]^{N\circ T}$ are the same. Moreover, we will see that they satisfy the Jacobi identity.

\begin{pro}\label{pro:S-Nijenhuis operator}
 Let $(T,S,N)$ be a Kupershmidt-(dual-)Nijenhuis structure. Then $S$ is a Nijenhuis operator on the sub-adjacent Lie algebra $(V,[-,-]^T)$. Thus, the brackets  ${[-,-]}^T_S,$ $\{-,-\}_{\hat{\varrho}}^T~(\{-,-\}_{\tilde{\varrho}}^T)$ and $[-,-]^{N\circ T }$ are all Lie brackets.
\end{pro}
\pf For the Kupershmidt-Nijenhuis structure $(T,S,N)$, by \eqref{eq:Nijpair} and substituting $x$ by $T(u)$, we get
    \begin{eqnarray*}
     0&=&\rho(NT(u))S(v)-S(\rho(NT(u))(v)+S\rho(T(u))S(v)-S\rho(T(u))(v))\\
     &=&\rho(TS(u))S(v)-S(\rho(TS(u))(v)+S\rho(Tu)S(v)-S\rho(Tu)(v))\\
     &=&S(u)\star^TS(v)-S(S(u)\star^Tv+u\star^TS(v)-S(u\star^Tv)),
     \end{eqnarray*}
     which implies that $S$ is a Nijenhuis operator on the pre-Lie algebra $(V,\star^T)$ (\cite{WBLS}). Thus $S$ is a Nijenhuis operator on the sub-adjacent Lie algebra $(V,[-,-]^T)$.

For the Kupershmidt-dual-Nijenhuis structure $(T,S,N)$, the proof is not direct. In fact, by the relation $[u,v]^{T\circ S}={[u,v]}^{T}_{S}$, one has
\begin{eqnarray}
 S(\rho({T\circ S(u)})(v))-S(\rho({T(v)})(S(u)))&=&\rho({T\circ S(u)})(S(v))-\rho({T(v)})(S^2(u));\label{eq:pn11}\\
 S^2(\rho({T(u)})(v))-S^2(\rho({T(v)})(u))&=& S(\rho({T(u)})(S(v)))-S(\rho({T(v)})(S(u))).
\end{eqnarray}
By the condition \eqref{eq:coNijpair}, we have
\begin{equation}\label{eq:pn22}
 \rho({ T(v)})(S^2(u))-\rho({T\circ S(v)})(S(u))=S(\rho({ T(v)})(S(u)))-S(\rho({T\circ S(v)})(u)).
\end{equation}
By \eqref{eq:pn11}-\eqref{eq:pn22}, we have
\begin{eqnarray*}
  &&[S(u),S(v)]^{T}- S([u,v]^{T}_{S})\\
  &=&\rho({T\circ S(u)})(S(v))-\rho({T\circ S(v)})(S(u))+S^2(\rho({T(u)})(v))-S^2(\rho({T(v)})(u))\\
  &&-S(\rho({T(u)})(S(v)))+S(\rho({T\circ S(v)})(u))-S(\rho({T\circ S(u)})(v))+S(\rho({T(v)})(S(u)))\\
&=&\rho({T\circ S(u)})(S(v))-\rho({T\circ S(v)})(S(u))-S^2(\rho({T(v)})(u))+S^2(\rho({T(u)})(v))\\
  &&-S(\rho({T(u)})(S(v)))+S(\rho({T\circ S(v)})(u))-\rho({T\circ S(u)})(S(v))+S(\rho({T(v)})(S^2(u)))\\
&=&\rho({T\circ S(u)})(S(v))-\rho({T\circ S(v)})(S(u))+ S(\rho({T(u)})(S(v)))-\rho({T(v)})(S(u))\\
  &&-S(\rho({T(u)})(S(v)))+S(\rho({T\circ S(v)})(u))-\rho({T\circ S(u)})(S(v))+\rho({T(v)})(S^2(u))\\
  &=&\rho({T(v)})(S^2(u))-\rho({T\circ S(v)})(S(u))-S(\rho({T(v)})(S(u)))+S(\rho({T\circ S(v)})(u))\\
  &=&0.
\end{eqnarray*}
Thus $S$ is a Nijenhuis operator on the Lie algebra $(V,[-,-]^{T})$.\qed
\emptycomment{By the proof of the above proposition, we can obtain a more strong conclusion for a Kupershmidt-Nijenhuis structure.
\begin{cor}
 Let $(T,S,N)$ be a Kupershmidt-Nijenhuis structure. Then $S$ is a Nijenhuis operator on the pre-Lie algebra $(V,\star^T)$.
\end{cor}}

\begin{thm}\label{pro:TS1}
 Let $(T,S,N)$ be a   Kupershmidt-(dual-)Nijenhuis structure. Then we have
  \begin{itemize}
\item[$\rm(i)$] $T$ is  a Kupershmidt operator on the deformed Lie algebra $(\g,[-,-]_N)$ with respect to the representation $(V; \hat{\varrho})$ $((V; \tilde{\varrho}))$;
 \item[$\rm(ii)$] $N\circ T$ is a Kupershmidt operator on the Lie algebra $(\g,[-,-]_{\g})$ with respect to the representation  $(V;\rho)$.
  \end{itemize}

\end{thm}
\pf
 We only prove the theorem for the Kupershmidt-Nijenhuis structure. The other one can be proved similarly.

(i)   Since $T$ is a Kupershmidt operator on the Lie algebra $(\g,[-,-]_\g)$ with respect to the representation $(V;\rho)$ and $T\circ S=N\circ T$, we have
\begin{eqnarray*}
T\{u,v\}^T_{\hat{\rho}}&=&T({[u,v]}^{T}_ {S})=T([S(u),v]^T+[u,S(v)]^T-S[u,v]^T)\\
&=&[T\circ S(u),T(v)]_\g+[T(u),T\circ S(v)]_\g-T\circ S[u,v]^T\\
&=&[N\circ T(u),T(v)]_\g+[T(u),N\circ T(v)]_\g-N\circ T[u,v]^T\\
&=&[T(u),T(v)]_N.
\end{eqnarray*}
Thus, $T$ is  a Kupershmidt operator on the deformed Lie algebra $(\g,[-,-]_N)$ with respect to the representation $(V; \hat{\varrho})$.

(ii) By \eqref{eq:TN2}, we have
\begin{eqnarray*}
 N\circ T([u,v]^{N\circ T})&=&N \circ T([u,v]^{T}_ {S})=N[T(u),T(v)]_N=[N\circ T(u),N\circ T(v)]_\g,
\end{eqnarray*}
which implies that $N\circ T$ is a Kupershmidt operator on the Lie algebra $(\g,[-,-]_{\g})$ with respect to the representation  $(V;\rho)$.
\qed\vspace{3mm}

The following theorem demonstrates that the Kupershmidt-Nijenhuis operator can give a Kupershmidt-dual-Nijenhuis operator with a
condition.
\begin{thm}\label{thm:dual-Nijenhuis and Nijenhuis}
 Let $(T,S,N)$ be a Kupershmidt-Nijenhuis structure. If $T$ is invertible, then $(T,S,N)$ is a Kupershmidt-dual-Nijenhuis structure.
\end{thm}
\pf We only need to prove that the Nijenhuis pair $(S,N)$ is also a dual-Nijenhuis pair. By \eqref{eq:TN2}, we have
\begin{eqnarray*}
{[u,v]}^{T}_{S}-{[u,v]}^{T\circ S}=\rho(T(u))(S(v))-\rho(T(v))(S(u))-S\Big(\rho(T(u))(v)-\rho(T(v))(u)\Big),
\end{eqnarray*}
which implies that
\begin{equation}\label{eq:TSN5}
\rho(T(u))(S(v))-\rho(T(v))(S(u))=S\Big(\rho(T(u))(v)-\rho(T(v))(u)\Big).
\end{equation}
Since $S$ is a Nijenhuis operator on the Lie algebra $(V,[-,-]^T)$ and ${[u,v]}^{T}_{S}={[u,v]}^{T\circ S}$, we have
$$S([u,v]^{T\circ S})=[S(u),S(v)]^T,$$
which means that
\begin{eqnarray}\label{eq:TSN6}
  S\Big(\rho(T\circ S(u))(v)-\rho(T\circ S(v))(u)\Big)=\rho(T\circ S(u))(S(v))-\rho(T\circ S(v))(S(u)).
\end{eqnarray}
By \eqref{eq:TSN5}, we have
\begin{eqnarray*}
 S\big(\rho(T\circ S(u))(v)\big)-\rho(T\circ S(u))(S(v))=S\big(\rho(T(v))(S(u))\big)-\rho(T(v))(S^2(u)).
\end{eqnarray*}
Thus \eqref{eq:TSN6} implies that
\begin{eqnarray*}
 0&=&S\big(\rho(T(v))(S(u))\big)-\rho(T(v))(S^2(u))- S\big(\rho(T\circ S(v))(u)\big)+\rho(T\circ S(v))(S(u))\\
 &=&S\big(\rho(T(v))(S(u))\big)-\rho(T(v))(S^2(u))- S\big(\rho(N\circ T(v))(u)\big)+\rho(N\circ T(v))(S(u)).
\end{eqnarray*}
Since $T$ is invertible and let $x=T(v)$, we have
$$S\big(\rho(x)(S(u))\big)-\rho(x)(S^2(u))- S\big(\rho(N(x))(u)\big)+\rho(N(x))(S(u))=0.$$
Thus the Nijenhuis pair $(S,N)$ is a dual-Nijenhuis pair. We finish the proof.\qed

\section{Hierarchy of Kupershmidt operators}

\subsection{Compatible Kupershmidt operators on Lie algebras}

\begin{defi}
  Let $T_1,T_2: V\longrightarrow \g$ be two
Kupershmidt operators on a Lie algebra $(\g,[-,-]_\g)$ with respect to a
representation $(V;\rho)$. If for all
$k_1,k_2$, $k_1T_1+k_2T_2$ is still a Kupershmidt operator, then $T_1$ and $T_2$ are called {\bf
compatible}.
\end{defi}

\begin{pro}\label{pro:compatible}
Let $T_1,T_2: V\longrightarrow \g$ be two Kupershmidt operators
on a Lie algebra $(\g,[-,-]_\g)$ with respect to a
representation $(V;\rho)$. Then $T_1$ and $T_2$ are compatible
if and only if the following equation holds:
\begin{eqnarray}
\nonumber [T_1(u), T_2(v)]_\g+ [T_2(u),
 T_1(v)]_\g&=&T_1\Big(\rho(T_2(u))(v)-\rho(T_2(v))(u)\Big)\\
 &&+T_2\Big(\rho(T_1(u))(v)-\rho(T_1(v))(u)\Big),\quad \forall u,v\in V.\label{eq:CN1}
\end{eqnarray}
\end{pro}

\pf It follows from a direct computation.\qed\vspace{3mm}

Using a Kupershmidt operator and a Nijenhuis operator, we can construct a pair of compatible Kupershmidt operators .

\begin{pro}\label{pro:NT}
Let $T: V\longrightarrow \g$ be a Kupershmidt operator on a
Lie algebra $(\g,[-,-]_\g)$ with respect to a representation
$(V;\rho)$ and $N$ a Nijenhuis operator on $(\g,[-,-]_\g)$.
Then $N\circ T$ is a Kupershmidt operator on the Lie algebra $(\g,[-,-]_\g)$ with respect to the representation $(V;\rho)$
if and only if for all $u,v\in V$, the following equation holds:
\begin{eqnarray}
\nonumber &&N\Big([NT(u), T(v)]_\g+[T(u), NT(v)]_\g\Big)\\
&=&N\Big(T\big(\rho(NT(u))(v)-\rho(NT(v))(u)\big)+NT\big(\rho(T(u))(v)-\rho(T(v))(u)\big)\Big).\label{eq:ON}
\end{eqnarray}
In this case, if in addition $N$ is invertible, then $T$ and $N\circ T$
are compatible. More explicitly, for any Kupershmidt operator $T$,
if there exists an invertible Nijenhuis operator $N$ such that $N\circ T$
is also a Kupershmidt operator, then $T$ and $N\circ T$ are compatible.
\end{pro}

\pf    Since $N$ is a Nijenhuis operator,
we have
$$[NT(u), NT(v)]_\g=N\Big([NT(u),T(v)]_\g+[T(u),
NT(v)]_\g\Big)-N^2([T(u), T(v)]_\g).$$ Note that
$$[T(u), T(v)]_\g=T\Big(\rho(T(u))(v)-\rho(T(v))(u)\Big).$$
Then
$$[NT(u), NT(v)]_\g=NT\Big(\rho(NT(u))(v)-\rho(NT(v))(u)\Big)$$
if and only if  (\ref{eq:ON}) holds.

If $N\circ T$ is a Kupershmidt operator and $N$ is invertible, then we have
$$[NT(u), T(v)]_\g+[T(u), NT(v)]_\g=
T\big(\rho(NT(u))(v)-\rho(NT(v))(u)\big)+NT\big(\rho(T(u))(v)-\rho(T(v))(u)\big),$$
which is exactly the condition that $N\circ T$ and $T$ are compatible. \qed\vspace{3mm}

A pair of compatible Kupershmidt operators can also give rise to a Nijenhuis operator under some conditions.

\begin{pro}\label{pro:TTN}
Let $T_1,T_2: V\longrightarrow \g$ be two Kupershmidt operators  on
a Lie algebra $(\g,[-,-]_\g)$ with respect to a representation
$(V;\rho)$. Suppose that $T_2$ is invertible. If $T_1$ and $T_2$
are compatible, then $N=T_1\circ T_2^{-1}$ is a Nijenhuis operator on the Lie algebra $(\g,[-,-]_\g)$.
\end{pro}

\pf For all $x,y\in \g$, there exist $u,v\in V$ such that
$T_2(u)=x, T_2(v)=y$. Hence $N=T_1\circ T_2^{-1}$ is a Nijenhuis operator
if and only if the following equation holds:
$$[NT_2(u), NT_2(v)]_\g=N([NT_2(u), T_2(v)]_\g+[T_2(u),
NT_2(v)]_\g)-N^2([T_2(u), T_2(v)]_\g).$$ Since $T_1=N\circ T_2$ is an
Kupershmidt operator, the left hand side of the above equation is
$$NT_2(\rho(NT_2(u))(v)-\rho(NT_2(v))(u)).$$
Since $T_2$ is a Kupershmidt operator which is compatible with
$T_1=N\circ T_2$, we have
\begin{eqnarray*}
 &&[NT_2(u),T_2(v)]_\g+[T_2(u), NT_2(v)]_\g\\
 &=&
T_2(\rho(NT_2(u))(v)-\rho(NT_2(v))(u))+NT_2(\rho(T_2(u))(v)-\rho(T_2(v))(u))\\
&=&T_2(\rho(NT_2(u))(v)-\rho(NT_2(v))(u))+N([T_2(u),T_2(v)]_\g).
\end{eqnarray*}
Let $N$ act on both sides, we get the conclusion. \qed\vspace{3mm}

By Proposition \ref{pro:NT} and \ref{pro:TTN}, we have
\begin{cor}\label{cor:NTN}
Let $T_1,T_2: V\longrightarrow \g$ be two Kupershmidt operators  on
a Lie algebra $(\g,[-,-]_\g)$ with respect to a representation
$(V;\rho)$. Suppose that $T_1$ and $T_2$ are invertible. Then
$T_1$ and $T_2$ are compatible if and only if $N=T_1\circ T_2^{-1}$ is a
Nijenhuis operator.
\end{cor}

In particular, as a direct application, we have the following conclusion.

\begin{cor} Let $(\g,[-,-]_\g)$ be a Lie algebra. Suppose that
$\huaR_1$ and $\huaR_2$ are two invertible Rota-Baxter operators. Then $\huaR_1$ and $\huaR_2$ are compatible in the sense that any
linear combination of $\huaR_1$ and $\huaR_2$ is still a Rota-Baxter
operator if and only if $N=\huaR_1\circ \huaR_2^{-1}$ is a
Nijenhuis operator.
\end{cor}

\subsection{Hierarchy of Kupershmidt  operators }
In the following, first we construct compatible Kupershmidt operators  from Kupershmidt-(dual-)Nijenhuis structures. Given a Kupershmidt-(dual-)Nijenhuis structure $(T,S,N)$, by Theorem \ref{pro:TS1}, $T$ and $T\circ S$ are Kupershmidt operators. In fact, they are compatible.
\begin{pro}\label{pro:TS2}
 Let $(T,S,N)$ be a Kupershmidt-(dual-)Nijenhuis structure. Then $T$ and $T\circ S$ are compatible Kupershmidt operators.
\end{pro}
\pf  We only prove the conclusion for the Kupershmidt-Nijenhuis structure. The other one can be proved similarly.
It is sufficient to prove that $T+T\circ S$ is a Kupershmidt operator.
It is obvious that
$$[u,v]^{T+T\circ S}=[u,v]^T+[u,v]^{T\circ S}=[u,v]^T+{[u,v]}^T_S.$$
Thus, we have
\begin{eqnarray*}
  &&(T+T\circ S)([u,v]^{T+T\circ S})\\
  &=&T([u,v]^T)+T\circ S([u,v]^T_S)+T\circ S([u,v]^T)+ T([u,v]^T_S)\\
  &=&T([u,v]^T)+T\circ S([u,v]^T_S)+T\circ S([u,v]^T)\\
  &&+T([S(u),v]^T+[u,S(v)]^T-S[u,v]^T)\\
  &=&T([u,v]^T)+T\circ S([u,v]^T_S)+T([S(u),v]^T+[u,S(v)]^T)\\
  &=&[T(u),T(v)]_\g+[T\circ S(u),T\circ S(v)]_\g+[T\circ S(u),T(v)]_\g+[T(u),T\circ S(v)]_\g\\
  &=&[(T+T\circ S)(u),(T+T\circ S)(v)]_\g,
\end{eqnarray*}
which means that $T+T\circ S$ is a Kupershmidt operator.\qed

\begin{lem}
 Let $(T,S,N)$ be a Kupershmidt-(dual-)Nijenhuis structure. Then for all $k,i\in\Nat$, we have
\begin{eqnarray}
   \label{eq:TSN3}T_k[u,v]^T_{S^{k+i}}=[ T_k(u), T_k(v)]_{N^i}.
\end{eqnarray}
\end{lem}
\pf
Since $T$ is a Kupershmidt operator and $T\circ S=N\circ T$, we have
\begin{eqnarray}
\nonumber T([u,v]^T_{S^i})&=&T\big([S^i(u), v]^T+[u, S^i(v)]^T-S^{i}([u, v]^T)\big)\\
\nonumber&=&[N^i(T (u)), T(v)]_\g+[T(u), N^i(T(v))]_\g-N^i([T(u), T(v)]_\g)\\
\label{eq:TSN2}&=& [T(u), T(v)]_{N^i}.
\end{eqnarray}
Since $S$ is a Nijenhuis operator on the associative algebra  $(V,\cdot^T)$, we have
\begin{equation}\label{eq:t1}
  S^k([u,v]^{T}_{S^{k+i}})=[S^k(u), S^k(v)]^{T}_{S^i}.
\end{equation}
Then by \eqref{eq:TSN2} and \eqref{eq:t1}, we have
\begin{eqnarray*}
 T_k([u,v]^{T}_{S^{k+i}}=T\circ S^k([u,v]^{T}_{S^{k+i}})=T([S^k(u),S^k(v)]^{T}_{S^i})= [T(S^k(u)),T(S^k(v))]_{N^i}.
\end{eqnarray*}
The proof is finished. \qed\vspace{3mm}

The proof of the following lemma is similar to the proof of Proposition 5.1 in \cite{Kosmann1}.
\begin{lem}
Let $(T,S,N)$ be a Kupershmidt-(dual-)Nijenhuis structure. Then for all $k,i\in\Nat$ such that $i\leq k$,
\begin{equation}\label{eq:TSN4}
  [u,v]^{T_k}= [u,v]^{T}_{S^k}=S^{k-i}[u,v]^{T_i},
\end{equation}
where $T_k=T\circ S^k=N^k\circ T$ and set $T_0=T$.
\end{lem}

\begin{pro}\label{pro:hierarchy}
Let $(T,S,N)$ be a Kupershmidt-(dual-)Nijenhuis structure with respect to the representation $(V;\rho)$. Then all $T_k=N^k\circ T$ are Kupershmidt operators with respect to the representation $(V;\rho)$ and for all $k,l\in\Nat$, $T_k$ and $T_l$ are compatible.
\end{pro}
\pf  We only prove the conclusion for the Kupershmidt-Nijenhuis structure. The other one can be proved similarly.

By  \eqref{eq:TSN3} and \eqref{eq:TSN4} with $i=0$, we have
\begin{eqnarray*}
 T_k[u,v]^{T_{k}}=[ T_k(u), T_k(v)]_\g,
\end{eqnarray*}
which implies that $T_k$ is a Kupershmidt operator associated the representation $(V;\rho)$.

For the second conclusion, we need  to prove that $T^k+T^{k+i}$ is a Kupershmidt operator for all $k,i\in\Nat$.
By  \eqref{eq:TSN4}, we have
$$[u,v]^{T_k+T_{k+i}}=[u,v]^{T_k}+[u,v]^{T_{k+i}}=[u,v]^{T_k}+[u,v]^{T_k}_{S^i}.$$
Thus, we have
\begin{eqnarray*}
  &&(T_k+T_{k+i})([u,v]^{T_k+T_{k+i}})\\
  &=&T_k([u,v]^{T_k})+T_k([u,v]^{T_k}_{S^i})+T_{k+i}([u,v]^{T_k})+T_{k+i}([u,v]^{T_k}_{S^i})\\
  &=&T_k([u,v]^{T_k})+T_{k+i}([u,v]^{T_k}_{S^i})+T_{k+i}([u,v]^{T_k}_{S^i})\\
  &&+T_k([S^i(u),v]^{T_k}+[u,S^i(v)]^{T_k}-S^i[u,v]^{T_k})\\
  &=&T_k([u,v]^{T_k})+T_{k+i}([u,v]^{T_k}_{S^i})+T^k([S^i(u),v]^{T_k})+T^k([u,S^i(v)]^{T_k})\\
  &=&[T_k(u),T_k(v)]_\g+[T_{k+i}(u),T_{k+i}(v)]_\g+[T_{k+i}(u),T_k(v)]_\g+[T_k(u),T_{k+i}(v)]_\g\\
  &=&[(T_k+T_{k+i})(u),(T_k+T_{k+i})(v)]_\g.
\end{eqnarray*}
Thus $T^k+T^{k+i}$ is a Kupershmidt operator. We finish the proof.\qed\vspace{3mm}

\emptycomment{

The Kupershmidt operator can be related to the solution of classical Yang-Baxter equation.
\begin{lem}{\rm\cite{Bai1}}
  $\pi^\sharp=T-T^*$ is an $r$-matrix in the Lie algebra $\g\ltimes_{\rho^*}V^*$ if and only if $T$ is a Kupershmidt operator associated to a representation $(V;\rho)$, where $T^*:\g^*\longrightarrow V^*$ is given by
  $$\langle T^*(\xi),u\rangle=\langle \xi,T(u)\rangle,\quad\forall\xi\in\g^*,u\in V.$$
\end{lem}

As a corollary of \ref{pro:hierarchy}, we have
\begin{cor}
 Let $(T,S,N)$ be a Kupershmidt-(dual-)Nijenhuis structure associated to the representation $(V;\rho)$. Then all $\pi^\sharp_k=N^k\circ T-T^*\circ (N^*)^k$ are $r$-matrix in the Lie algebra $\g\ltimes_{\rho^*}V^*$ and for all $k,l\in\Nat$, $\pi^\sharp_k$ and $\pi^\sharp_l$ are compatible.
\end{cor}
}

Compatible Kupershmidt operators  can give rise to Kupershmidt-dual-Nijenhuis structures.
\begin{pro}\label{pro:ComptoNS}
Let $T,T_1: V\longrightarrow \g$ be two Kupershmidt operators  on
a Lie algebra $(\g,[-,-]_\g)$ with respect to a representation
$(V;\rho)$. Suppose that $T$ is invertible. If
$T$ and $T_1$ are compatible, then
 \begin{itemize}
\item[$\rm(i)$]$(T,S=T^{-1}\circ T_1,N=T_1\circ T^{-1})$ is a Kupershmidt-dual-Nijenhuis structure;
\item[$\rm(ii)$]$(T_1,S=T^{-1}\circ T_1,N=T_1\circ T^{-1})$ is a Kupershmidt-dual-Nijenhuis structure.
\end{itemize}
\end{pro}
\pf (i) The proof of $(N,S)$ being a dual-Nijenhuis pair   is similar to the proof of Theorem \ref{thm:dual-Nijenhuis and Nijenhuis}. We omit the details.
It is obvious that $T\circ S=N\circ T$. Thus we only need to prove that the compatibility condition \eqref{eq:TN2} holds.
By the compatibility condition of $T$ and $T_1$ and Proposition \ref{pro:TTN}, $N=T_1\circ T^{-1}$ is a Nijenhuis operator on the Lie algebra $\g$. By Proposition \ref{pro:compatible}, we also have
\begin{eqnarray*}
 [T(u), T_1(v)]_\g+ [T_1(u),
 T(v)]_\g&=&T\Big(\rho(T_1(u))(v)-\rho(T_1(v))(u)\Big)\\
 &&+T_1\Big(\rho(T(u))(v)-\rho(T(v))(u)\Big),\quad \forall u,v\in V.
\end{eqnarray*}
Substituting $T_1$ with $T\circ S$, then we have
\begin{eqnarray}
 \nonumber[T(u), T\circ S(v)]_\g+ [T\circ S(u),
 T(v)]_\g&=&T\Big(\rho(T\circ S(u))(v)-\rho(T\circ S(v))(u)\Big)\\
 &&+T\circ S\Big(\rho(T(u))(v)-\rho(T(v))(u)\Big).\label{eq:Comp1}
\end{eqnarray}
Since $T$ is a Kupershmidt operator, we have
\begin{eqnarray*}
 [T(u), T\circ S(v)]_\g+ [T\circ S(u), T(v)]_\g&=&T\Big(\rho(T(u))(S(v))-\rho(T\circ S(v))(u)\\
 &&+\rho(T\circ S(u))(v)-\rho(T(v))(S(u))\Big).
\end{eqnarray*}
Since $T$ is invertible, \eqref{eq:Comp1} is equivalent to
\begin{eqnarray}\label{eq:ON3}
  S\Big(\rho(T(u))(v)-\rho(T(v))(u)\Big)=\rho(T(u))(S(v))-\rho(T(v))(S(u)).
\end{eqnarray}
On the other hand, we have
\begin{eqnarray*}
{[u,v]}^{T}_{S}-{[u,v]}^{T\circ S}=\rho(T(u))(S(v))-\rho(T(v))(S(u))-S\Big(\rho(T(u))(v)-\rho(T(v))(u)\Big).
\end{eqnarray*}
Thus, \eqref{eq:ON3} implies that $ {[u,v]}^{T}_{S}={[u,v]}^{T\circ S}$.
Therefore, $(T,S=T^{-1}\circ T_1,N=T_1\circ T^{-1})$ is a Kupershmidt-dual-Nijenhuis structure.

\emptycomment{
Furthermore, since $T$ and $T\circ S$ are Kupershmidt operators, thus
$$T\circ S([u,v]^{T\circ S})=[T\circ S(u),T\circ S(v)]_\g=T([S(u),S(v)]^T).$$
As $T$ is invertible, we have
$$S([u,v]^{T\circ S})=[S(u),S(v)]^T.$$
}

(ii) By direct calculation, we have
\begin{eqnarray*}
 &&{[u,v]}^{T_1}_{S}- {[u,v]}^{T_1\circ S}\\&=&\rho(T_1(u))(S(v))-\rho(T_1(v))(S(u))-S\Big(\rho(T_1(u))(v)-\rho(T_1(v))(u)\Big)\\
  &=&\rho(T\circ S(u))(S(v))-\rho(T\circ S(v))(S(u))-S\Big(\rho(T\circ S(u))(v)-\rho(T\circ S(v))(u)\Big)\\
  &=&[S(u),S(v)]^T-S[u,v]^{T\circ S}=0.
\end{eqnarray*}
Thus, $(T_1,S=T^{-1}\circ T_1,N=T_1\circ T^{-1})$ is also a Kupershmidt-dual-Nijenhuis structure.
 \qed

\section{Rota-Baxter-Nijenhuis structures and  $r$-matrix-Nijenhuis structures}
In the following, we first recall the definition of an $r$-$n$ structure on a Lie algebra $\g$, which is the infinitesimal of a right-invariant Poisson-Nijenhuis structure on the  Lie group $G$ integrating the Lie algebra $\g$ (\cite{rn}).  We call such a structure an $r$-matrix-Nijenhuis structure.
\begin{defi}\label{defi:rmnij}
Let $\pi$ be an $r$-matrix and $N:\g\longrightarrow \g$ a Nijenhuis operator on a Lie algebra $(\g,[-,-]_\g)$. A pair $(\pi,N)$ is a {\bf $r$-matrix-Nijenhuis structure} on the Lie algebra $(\g,[-,-]_\g)$ if for $x,y\in\g$ and $\alpha,\beta\in\g^*$, they satisfy
\begin{eqnarray}
 \label{eq:rmn1} N\circ \pi^\sharp&=&\pi^\sharp\circ N^*,\\
 \label{eq:rmn2} {[\alpha,\beta]}^{N\circ \pi^\sharp}&=&{[\alpha,\beta]}_{N^*}^{\pi^\sharp},
\end{eqnarray}
where
$\pi^\sharp:\g\longrightarrow \g$ is a linear operator induced by $\langle \pi^\sharp(\alpha),\beta\rangle=\pi(\alpha,\beta)$
, \eqref{eq:rmn2} is given by \eqref{eq:TN2} with $S=N^*$, $T=\pi^\sharp$ and the representation $\rho=\ad^*$.
\end{defi}
It is obvious that the triple $(\pi,S=N^*,N)$ is a Kupershmidt-dual-Nijenhuis structure on the Lie algebra $\g$ with respect to the representation $(\g^*;\ad^*)$.

Similar to the $r$-matrix-Nijenhuis structure, we give the definition of Rota-Baxter-Nijenhuis structure on a Lie algebra.
\begin{defi}
Let $(\g,[\cdot,\cdot]_\g)$ be a Lie algebra. Let $\huaR:\g\longrightarrow\g$ be a Rota-Baxter operator and $N:\g\longrightarrow \g$  a Nijenhuis operator on the Lie algebra $\g$. A pair $(\huaR,N)$ is a {\bf Rota-Baxter-Nijenhuis structure} on the Lie algebra $\g$  if for $x,y\in\g$, they satisfy
\begin{eqnarray}
 \label{eq:YBTS1}N\circ \huaR&=&\huaR\circ N,\\
 \label{eq:YBTS2}[x,y]^{N \circ\huaR}&=&[x,y]_{N}^{\huaR}.
\end{eqnarray}
where \eqref{eq:YBTS2} is given by \eqref{eq:TN2} with $T=\huaR$ and the representation $\rho=\ad$.
\end{defi}

It is obvious that if $(R,N)$ is Rota-Baxter-Nijenhuis structure, then the triple $(R,S=N,N)$ is a Kupershmidt-Nijenhuis structure on the Lie algebra $\g$ with respect to the representation $(\g;\ad)$.

In the following, we study the relation between Rota-Baxter-Nijenhuis structure and $r$-matrix-Nijenhuis structure. First we recall some notions which was given in the articles by Semonov-Tian-Shansky \cite{Semonov-Tian-Shansky} or Kosmann-Schwarzbach \cite{Kosmann1}.

Let $\g$ be a Lie algebra with an $\ad$-invariant, non-degenerate, symmetric bilinear form $B\in\g\otimes\g$. Then $B$ induces a bijective linear map $B^\sharp:\g^*\longrightarrow\g$ given by
\begin{equation}
  \langle B^\sharp(\alpha),\beta\rangle=B(\alpha,\beta),\quad\forall~\alpha,\beta\in\g^*.
\end{equation}
By the $\ad$-invariance of $B$, we have
\begin{equation}\label{eq:adinv}
  B^\sharp(\ad^*_x\alpha)=\ad_x(B^\sharp(\alpha)),\quad\forall~x\in\g,\alpha\in\g^*.
\end{equation}
A {\bf skew-symmetric endomorphism of $(\g,B)$} is a linear map $\huaR$ from $\g$ to $\g$ such that $\huaR\circ B^\sharp:\g^*\longrightarrow\g$ is skew-symmetric.

The following theorem demonstrates the relation between Rota-Baxter-Nijenhuis structure and $r$-matrix-Nijenhuis structure.
\begin{thm}\label{thm:rmatrix-RBN}
 Let $\huaR$ be a skew-symmetric endomorphism of $(\g,B)$, $N:\g\longrightarrow\g$ a Nijenhuis operator, and set $\pi^\sharp=\huaR\circ B^\sharp$. Assume that $B$ and $N$ are compatible, i.e.
 \begin{equation}\label{eq:ad-bilinear}
 B^\sharp\circ N^*=N\circ B^\sharp.
 \end{equation}
If $(\huaR,N)$ is a Rota-Baxter-Nijenhuis structure on the Lie algebra $\g$, then $(\pi,N)$ is an $r$-matrix-Nijenhuis structure on the Lie algebra $\g$. Conversely, let $(\pi,N)$ be a r-matrix-Nijenhuis structure on the Lie algebra $\g$ with an $\ad$-invariant, non-degenerate, symmetric bilinear form $B$. Then $(\huaR=\pi^\sharp\circ(B^\sharp)^{-1} ,N)$ is a Rota-Baxter-Nijenhuis structure on the Lie algebra $\g$.
\end{thm}
\pf By Semonov-Tian-Shansky's conclusion, $\pi$ is an $r$-matrix. By \eqref{eq:YBTS1} and \eqref{eq:ad-bilinear}, it is obvious that $N\circ \pi^\sharp=\pi^\sharp\circ N^*$.

Let $\alpha=(B^\sharp)^{-1}(x),\beta=(B^\sharp)^{-1}(y)$, by \eqref{eq:adinv}, we have
\begin{eqnarray*}
   {[\alpha,\beta]}^{\pi^\sharp}&=&\ad^*_{ \pi^\sharp((B^\sharp)^{-1}(x))}(B^\sharp)^{-1}(y)-\ad^*_{ \pi^\sharp((B^\sharp)^{-1}(y))}(B^\sharp)^{-1}(x)\\
   &=&\ad^*_{ \huaR(x)}(B^\sharp)^{-1}(y)-\ad^*_{ \huaR(y)}(B^\sharp)^{-1}(x)\\
   &=&(B^\sharp)^{-1}([\huaR(x),y]_\g-[\huaR(y),x]_\g)=(B^\sharp)^{-1}([x,y]^{\huaR}),
\end{eqnarray*}
which implies that
\begin{equation}\label{eq:rmn3}
 {[\alpha,\beta]}^{\pi^\sharp}=(B^\sharp)^{-1}([x,y]^{\huaR}).
\end{equation}
Thus by \eqref{eq:ad-bilinear} and \eqref{eq:rmn3}, one has
\begin{eqnarray*}
  &&{[\alpha,\beta]}^{N\circ \pi^\sharp}-{[\alpha,\beta]}_{N^*}^{\pi^\sharp}={[\alpha,\beta]}^{N\circ \pi^\sharp}-\big([N^*(\alpha),\beta]^{\pi^\sharp}+[\alpha,N^*(\beta)]^{\pi^\sharp}-N^*[\alpha,\beta]^{\pi^\sharp}\big)\\
  &=&(B^\sharp)^{-1}([x,y]^{N\circ \huaR})-\big([N^*((B^\sharp)^{-1}(x)),(B^\sharp)^{-1}(y)]^{\pi^\sharp}+[(B^\sharp)^{-1}(x),N^*((B^\sharp)^{-1}(y))]^{\pi^\sharp}\\
  &&-N^*[(B^\sharp)^{-1}(x),(B^\sharp)^{-1}(y)]^{\pi^\sharp}\big)\\
  &=&(B^\sharp)^{-1}([x,y]^{N\circ \huaR})-\big([(B^\sharp)^{-1}\circ N(x),(B^\sharp)^{-1}(y)]^{\pi^\sharp}+[(B^\sharp)^{-1}(x),(B^\sharp)^{-1}\circ N(y))]^{\pi^\sharp}\\
  &&-N^*[(B^\sharp)^{-1}(x),(B^\sharp)^{-1}(y)]^{\pi^\sharp}\big)\\
  &=&(B^\sharp)^{-1}([x,y]^{N\circ \huaR})-(B^\sharp)^{-1}([N(x),y]_\g)-(B^\sharp)^{-1}[x,N(y)]_\g+N^*\circ (B^\sharp)^{-1}([x,y]_\g)\\
  &&-N^*[(B^\sharp)^{-1}(x),(B^\sharp)^{-1}(y)]^{\pi^\sharp}\big)\\
  &=&(B^\sharp)^{-1}([x,y]^{N\circ \huaR})-(B^\sharp)^{-1}\big([N(x),y]_\g+[x,N(y)]_\g-N([x,y]_\g)\big)\\
  &=&(B^\sharp)^{-1}([x,y]^{N\circ \huaR}-[x,y]^{\huaR}_N)=0,
\end{eqnarray*}
which implies that ${[\alpha,\beta]}^{N\circ \pi^\sharp}={[\alpha,\beta]}_{N^*}^{\pi^\sharp}$.

By a similar proof, the converse follows immediately. We finish the proof.\qed\vspace{3mm}

\noindent {\bf Acknowledgement:} This work was supported by NSFC (11471139,  11425104), NSF of Jilin Province(20170101050JC) and Nanhu Scholars Program for Young Scholars of XYNU.

\end{document}